\documentclass[preprintnumbers,amsmath,amssymb]{revtex4}
\def\be{\begin{equation}}
\def\ee{\end{equation}}
\def\bea{\begin{eqnarray}}
\def\eea{\end{eqnarray}}

\def\m{\mu}
\def\n{\nu}

\def\trr{\triangleright}
\def\p{\partial}
\def\a{\alpha}
\def\b{\beta}
\def\t{\theta}

\usepackage{graphicx}
\usepackage{dcolumn}
\usepackage{bm}
\usepackage{amsmath}
\usepackage{amssymb}
\usepackage{epsfig}

\begin{document}

\title{Scalar Field in Noncommutative Curved Space Time}
\author{Abolfazl\ Jafari}
\affiliation{Department of Physics, Faculty of Science,
Shahrekord University, P. O. Box 115, Shahrekord, Iran.\\
jafari-ab@sci.sku.ac.ir}
\date{\today }

\begin{abstract}
We study the issue of complex scalar field theories in noncommutative curved space time (NCCST)
with a new star-product.
In this paper, the equation of motion of
scalar field
and the canonical energy-momentum tensor of scalar field in static noncommutative curved space time (SNCCST)
will be found. The most important point is the assumption of the noncommutative parameter ($\theta$) be $x^{\m}$-independent,
the noncommutativity with time attended (TNC) and the metric tensor is time independent.
\end{abstract}


\maketitle


\section{Introduction}
Since Newton the concept of space time has gone through various
changes. All stages, however, had in common the notion of a continues linear
space. Today we formulate fundamental laws of physics, field theories, gauge field
theories, and the theory of gravity on differentiable manifolds.
That a change in the
concept of space for very short distances might be necessary was already anticipated
%in 1854
by Riemann.
% in his famous inaugural lecture.
There are indications today
that at very short distances we might have to go beyond differential manifolds. This
is only one of several arguments that we have to expect some changes in physics
for very short distances.
Other arguments are based on the singularity problem in
quantum field theory and the fact theory of gravity is non re-normalizable when
quantized. Why not try an algebraic concept of space time that could guide us to
changes in our present formulation of laws of physics? This is different from the
discovery of quantum mechanics.
There physics data forced us to introduce the concept
of noncommutativity.

The concepts of noncommutative subjects were born many years ago, where
the idea of noncommutative
coordinates is almost as old as quantum mechanics.
There are many approaches to noncommutative geometry and its use in physics. The operator algebra
and $\mathcal{C}^\star$-algebra one, the deformation quantization one, the quantum group one, and the matrix algebra
(fuzzy geometry) one.
Most of these approaches focus on
free or interacting QFTs on the Moyal-Weyl deformed or $\kappa$-deformed Minkowski
space time. The some physicists prefer to work in the noncommutative field theory, because
the NCFT has many ambiguous problems such as UV/IR divergence and causality\cite{wess,jab,nekrasov,jaf,panero}.

We briefly introduce the new star-product. By
deforming the ordinary Moyal-Wyle star-product ($\star$-product), we propose a new
star-product ($\trr$-product) which takes into consideration the missing terms cited above which generate
gravitational terms to the order $\t^2$. In Ref. \cite{khallili}, the authors have shown that we can
formulate of field theory on noncommutative curved space time
with replacing of operators variables.
% in following form:
We can replace the
noncommutative flat space time coordinates variables $[\hat{x}^\m\ ,\ \hat{x}^\n]=\imath\t^{\m\n}$
with noncommutative curved space time coordinates variables
$\hat{X}^\m=\hat{x}^\m+\imath^2\frac{\t^{ab}\t^{\a\b}}{2\sqrt{-\hat{g}}}\p_b \hat{R}^{\m}_{a\a\b}$
where $\hat{R}^\m_{b\a\b}$ stands for the Riemann's curvature tensor and $\sqrt{-\hat{g}}$ is determinant of metric, where
they are functions in noncommutative coordinates. This product is a nonassociative case, in contrast of the Moyal-Wyle $\star$-product but, with certain conditions,
would be an associative product.
For two any smooth functions, $A$ and $B$, we have
$\hat{A}(\hat{X})\hat{B}(\hat{X})=A(x).\ e^{
\frac{\imath}{2}\t^{\m\n}\overleftarrow{\p}_\m\otimes\overrightarrow{\p}_\n+\imath(\imath^2\frac{\t^{ab}\t^{\a\b}}{2\sqrt{-g}}\p_b R^{\m}_{a\a\b})
\p_\m\ (\Im_A\otimes\Im_B)}.\ B(x)$
where $\Im$ stands for the identity of vector spaces.
By considering of $\triangle^\m=\imath^2\frac{\t^{ab}\t^{\a\b}}{2\sqrt{-g}}\p_b \hat{R}^{\m}_{a\a\b}$ and
$\trr\equiv e^{\triangle x^\m \p_\m(\overleftarrow{\Im} \otimes\overrightarrow{\Im})
+\frac{\imath\theta^{\m\n}}{2}\p_\m\otimes \p_\n}$ we have
$(A\trr B)(\hat{x})=A\ e^{
\frac{\imath}{2}\t^{\m\n}\overleftarrow{\p}_\m\otimes\overrightarrow{\p}_\n+\imath\triangle^\m
\p_\m\ (\overleftarrow{\Im}\otimes\overrightarrow{\Im})}\ B$. When the $\triangle$ will be a $x^\m$-independent parameter,
we can write
$\int d^{(d-1)} x (A\trr B)(\hat{x})=\int d^{(d-1)} x (A(x)\star B(x)+\triangle^\m\p_\m(A(x)B(x))+0(\t^4))$
when all of functions to fall of faster than $\mid \vec{r}\mid^{-\frac{1}{2}(d-1)}$, so we can remove the last term and this implies that the new star product
will be an associative. In continue, we would like to introduce a new symbol $S_{\star}(A_1,A_2,..,A_n)$ where the $S_\star$ takes the total symmetrical structures
of $A_1$,...,$A_n$.

\section{construction of action and search of the equation of motion of fields}

We start by showing how to construct an action for scalar field (or vector scalar field) in the TNC with consideration of the metric tensor be time independent.
If one direct follows the general rule of transforming usual
theories in noncommutative ones by replacing product of fields by star product \cite{jab,nekrasov}
and we believe that these changes should be done on the Lagrangian density.
If the metric tensor will be time independent, so the Riemann's curvature tensor is time independent,
and the last term in $\trr-$product ($\frac{\t^{\a\b}\t^{\m\n}}{2\sqrt{-g}}\p_\b R^{\gamma}_{\a\m\n}\p_\gamma\ (\Im_A\otimes\Im_B)$) can be dropped out
so, we can write an associative noncommutative quantum field theory.
In
fact,
$s_{Cm}(\mathcal{L}_{Cm})\rightarrow s_{Nc}(\mathcal{L}^{Sym}_{Nc})\ $
%This says on
%Lagrangian should be done.
or
$$S=\int \textit{d}^d x\ \sqrt{-g}{\mathcal{L}_\trr}^{Sym}_{Nc}\ \ $$
%but for noncommutative with time attended, the $\trr$-product between the metric tensor and Lagrangian terms would be dropped out, because we can not research the
%dynamics of metric ( the momentum of metric ).
The classical Lagrangian density for scalar field in general space time is
\bea
\mathcal{L}=\frac{1}{2}\sqrt{-g}(g^{\m\n}\nabla_\m\Phi^{\dag}(x)\nabla_\n\Phi(x)-(m^2+\xi R(x))\Phi^2)
\eea
where $\Phi$ is the scalar field, $m$ is mass of field, $\nabla_\m$ is the covariant derivative where for the scalar field we have
$\nabla_\m\phi=\p_\m\phi$ and the coupling between the field and gravitational
field represented by the term $\xi R\Phi^2$ where $\xi$ is a numerical factor and $R(x)$ is the Ricci's scalar
curvature. But
because of the absence of quantum gravity certainly here, after we take $\xi=0$.
In the noncommutative curved space time with
$\trr$-product the action is more complicated
due to takeing into account symmetric ordering.

%First, the scalar field on the space time when $\triangle$ will be a constant.
We are
in particular interested in the deformation of the canonical action
\bea \label{3}
s&=&
\int \textit{d}^d x\ \ \frac{1}{2}\sqrt{-g}
(S_\trr(\nabla_\m\Phi,g^{\m\n},\nabla_\n\Phi)-
m^2\Phi^{\dag}\trr\Phi)\nonumber\\&&
\eea
we should  take the new covariant derivative $(\nabla_\m\star)$
because the curve space time is consideration
and we know $\nabla_\n\star A^\a=
\p_\n A^\a+\Gamma^{\a}_{\n \m}\star A^\m$.
If we have the spatial noncommutativity, the symmetric ordering can be choice
\bea \label{4}
S_\trr(\nabla_\m\Phi,g^{\m\n},\nabla_\n\Phi)
\eea
but in the case of TNC the metric tensor does not participate in the star-product and the earlier star product ($\trr$-product) will be an associative
(only if the $\triangle$ will be a constant).
By using these tools, we can deform the classical Lagrangian into TNC,
$2\mathcal{L}_{\star}=g^{\m\n}S_\star(\nabla_\m\Phi, \nabla_\n\Phi)-m^2 S_{\star}(\Phi,\Phi)$
so we can deform the classical action to the global expression
\bea \label{5}
s&=&
\int \textit{d}^d x\ \ \frac{1}{2}\sqrt{-g}(
(g^{\m\n}S_\star(\nabla_\m\Phi, \nabla_\n\Phi))-m^2\Phi^{a\dag}\trr\Phi_a)\nonumber\\&&
\eea
where
\bea
\Phi_a=\left(
         \begin{array}{c}
           \phi_1 \\
           \phi_2 \\
           .      \\
           .      \\
         \end{array}
       \right)
\eea

For the research of the equation of motion of field we can write the principle of the least action $\frac{\delta s}
{\delta\Phi_c(z)}=0$, namely,
\bea
\frac{\delta s}{\delta\Phi_c(z)}&=&\frac{\delta}{\delta\Phi_c(z)}
\int \textit{d}^d x
\frac{\sqrt{-g}}{2} (g^{\m\n}(\nabla_\m\Phi^{a\dag}\star\nabla_\n\Phi_a)-m^2 \Phi^{a\dag}\star\Phi_a)
\eea
but we know $\nabla_\m\star(A\star B_i)=\p_\m A\star B_i+A\star\nabla_\m\star B_i$ so we get to
\bea &&
\p_\n((\sqrt{-g}g^{\m\n})\star\nabla_\m\Phi^{c\dag})-
[\Gamma^{\n}_{\n\lambda}\ ,\ (\sqrt{-g}g^{\m\lambda})\star\nabla_\m\Phi^{c\dag}]_\star
+m^2 \sqrt{-g}\star\Phi^{c\dag}=0\nonumber\\&&
\eea
\bea &&
\p_\n(\nabla_\m\Phi_{c}\star(\sqrt{-g}g^{\m\n}))-
[\Gamma^{\n}_{\n\lambda}\ ,\ \nabla_\m\Phi_{c}\star(\sqrt{-g}g^{\m\lambda})]_\star
+m^2 \Phi_{c}\star\sqrt{-g}=0\nonumber\\&&
\eea

For real scalar field we get to
\bea &&
\p_\n\{\nabla_\m\Phi\ ,\ \sqrt{-g}g^{\m\n}\}_\star-
[\Gamma^{\n}_{\n\lambda}\ ,\ \{\nabla_\m\Phi\ ,\ \sqrt{-g}g^{\m\lambda}\}_\star]_\star
+m^2\{\Phi\ ,\ \sqrt{-g}\}_\star=0\nonumber\\&&
\eea

\section{The formal canonical energy-momentum tensor}

It is useful to study the derivative the different pieces of the action with respect to $g^{\m\n}$.
In view of the physical interpretations to be given later we introduce the new tensor $\mathbf{T}^{\m\n}$ via
$$\frac{-2\p}{\p\sqrt{-g}}\frac{\delta s}{\delta g_{\m\n}}=\mathbf{T}^{\m\n}$$
as the field symmetric energy-momentum tensor. Consider the action of field for $\t^{0i}\neq0$.
In this case, we can remove all $\star$'s related to metric tensor, because we do not search the momentum of
metric \cite{Neto2}.
However,
variation with respect to $g_{\m\n}$ for \ref{5} gives
\bea &&
\delta s=\int d^d x (\delta \sqrt{-g}\ \mathcal{L}_\star+
\sqrt{-g}\ \delta\mathcal{L}_\star),\nonumber\\&&
\eea
We first perform the variation of $\sqrt{-g}$ with respect to $\delta g_{\m\n}$. For this we write
$\delta \sqrt{-g}=-\frac{1}{2}\sqrt{-g} g_{\m\n}\delta g^{\m\n}$
\cite{kleinert}.
\bea
\delta s&=&\int d^dx (\frac{-1}{2}\sqrt{-g}g_{\m\n}\delta g^{\m\n}\
\mathcal{L}_\star+\sqrt{-g}\ \delta\mathcal{L}_\star),
\eea
consider now the variation of Lagrangian density, we get to
\bea
\mathbf{T}_{\eta\kappa}&=&2g_{\eta\kappa}\mathcal{L}_{\star}
-(\nabla_\eta\Phi^{a\dag}\star\nabla_\kappa\Phi_a+\nabla_\kappa\Phi^{a\dag}\star\nabla_\eta\Phi_a)\nonumber\\&&
\eea
and we see that $\mathbf{T}_{\m\n}=\mathbf{T}_{\n\m}$.

\textbf{Discussion}

We consider a new star-product in noncommutative geometry developed in a static curved space time
and we write an action for the complex scalar field theory in SNCCST and we study
the equation of motion of fields.
We are lead to the new equation of motion of field which it
is reduced to the equation of motion of scalar field in noncommutative flat space time when
the Riemann's curvature tensor will be zero and $\t$ be $x^\m$-independent. Additionally, we construct the
typical symmetric energy-momentum
tensor, from general way in static noncommutative curved space time.

\textbf{Acknowledgments}
The author thanks the Shahrekord University for support of this research.
\newline
%%%%%%%%%%%%%%%%%%%%%%%%%%%%%%%%%%%%%%%%%%%%%%%%%%


\begin{thebibliography}{99}
\bibitem{wess} Lect. Notes Phys. 774(Springer, Berlin Heidelberg 2009),
"Noncommutative Spacetimes: Symmetries in Noncommutative Geometry and Field Theory", DOI 10.1007/978-3
-540-89793-4

\bibitem{jab} M. Chaichian, P. Presnajder, M.M. Sheikh- Jabbari and A. Tureanu,
Eur.Phys.J. {\bf C29} (2003) 413-432, hep-th/0107055, and M. M.
Sheikh-Jabbari, J. High Energy Phys. {\bf 9906} (1999) 015,
hep-th/9903107, and I. F. Riad and M. M. Sheikh-Jabbari, J. High
Energy Phys. {\bf 0008} (2000) 045, hep-th/0008132, and H. Arfaei
and M. M. Sheikh-Jabbari, Nucl. Phys. {\bf B526} (1998) 278,
hep-th/9709054 and A. Jafari,``Recursive Relations For Processes With $n$ Photons
Of Noncommutative QED," Nucl. Phys. {\bf B783} (2007) 57-75, hep-th/0607141.

\bibitem{nekrasov} Nikita A. Nekrasov "
Trieste Lectures on solitons in Noncommutative Gauge Theories", hep-th/0011095.

\bibitem{jaf} Amir H. Fatollahi, Abolfazl Jafari, Eur. Phys. J. {\bf C46} (2006) 235-245.

\bibitem{panero} M. Panero "Numerical simulations of a non-commutative theory: the scalar model on the fuzzy sphere",
JHEP {\bf 0705} (2007) 082 and "Quantum Field Theory in a Non-Commutative Space: Theoretical Predictions and Numerical Results on the Fuzzy Sphere",
SIGMA {\bf 2} (2006) 081.

\bibitem{khallili} N. Mebaraki, F. Khallili, M. Boussahel and M. Haouchine, "
Modified Moyal-Weyl star product in a curved non commutative Space-Time",
Electron. J. Theor. Phys. {\bf 3} (2006) 37.

\bibitem{Neto2}R. Amorim, J. Barcelos-Neto,
"Remarks on the canonical quantization of noncommutative theories",
J. Phys. {\bf A34} (2001) 8851-8858 and "Noncommutative fields in curved space
", hep-th/0309145.

\bibitem{kleinert} Hagen Kleinert, "Path integral in Quantum Mechanics, Statistics,
Polymer Physics, and Financial Markets",  World Scientific Publishing Company,
(2009). and "Multivalued Fields: In Condensed Matter, Electromagnetism, and Gravitation
", World Scientific Publishing Company, (2008).

\end{thebibliography}
\end{document}